\documentclass[reprint,twocolumn]{revtex4}
\usepackage{amsfonts}
\usepackage{amsmath}
\usepackage{amssymb}
\usepackage{charter}
\usepackage{graphicx}
\usepackage{float}

\begin{document}

\title{Global quantum thermometry based on the optimal biased bound}
\author{Shoukang Chang$^{1}$}
\author{Wei Ye$^{2}$}
\thanks{Corresponding author. 71147@nchu.edu.cn}
\author{Xuan Rao$^{2}$}
\author{Huan Zhang$^{3}$}
\author{Liqing Huang$^{1}$}
\author{Mengmeng Luo$^{4}$}
\author{Yuetao Chen$^{1}$}
\author{Qiang Ma$^{1}$}
\author{Shaoyan Gao$^{1}$}
\thanks{Corresponding author. gaosy@xjtu.edu.cn}
\affiliation{$^{{\small 1}}$\textit{MOE Key Laboratory for Nonequilibrium Synthesis and
Modulation of Condensed Matter, Shaanxi Province Key Laboratory of Quantum
Information and Quantum Optoelectronic Devices, School of Physics, Xi'an
Jiaotong University, Xi'an 710049, People's Republic of China}\\
$^{{\small 2}}$\textit{School of Information Engineering, Nanchang Hangkong
University, Nanchang 330063, China}\\
$^{{\small 3}}${\small \ }\textit{School of Physics, Sun Yat-sen University,
Guangzhou 510275, China}\\
$^{4}$\textit{Department of Physics, Xi'an Jiaotong University City College,
Xi'an 710018, China}}

\begin{abstract}
Thermometry is a fundamental parameter estimation problem which is crucial
in the development process of natural sciences. One way to solve this
problem is to the extensive used local thermometry theory, which makes use
of the classical and quantum Cram\'{e}r-Rao bound as benchmarks of
thermometry precision. However, such a thermometry theory can only be used
for decreasing temperature fluctuations around a known temperature value and
hardly tackle the precision thermometry problem over a wide temperature
range. For this reason, we derive two basic bounds on thermometry precision
in the global setting and further show their thermometry performance by two
specific applications, i.e., noninteracting spin-$1/2$ gas and a general $N$%
-level thermal equilibrium quantum probe.

\end{abstract}

\maketitle

\section{Introduction}

Precision thermometry of a physical system is a non-trivial task in the
development of modern natural science and technology \cite{1,2,3,4}.
Especially, accurate temperature measurements are crucial for experiments at
the micro-and nanoscale and guaranteeing the performance of quantum
protocols \cite{5,6,7}; however, this remains a formidable task. On the one
hand, due to the fact that temperature is not a observable quantity but an
entropic quantity, which cannot be directly measured, one only utilize the
statistical behavior of some observable quantities to indirectly estimate
temperature \cite{8,9,10}. On the other hand, considering the fragility of
quantum systems, it is necessary to minimize interference while obtaining
maximum temperature measurement information, this undoubtedly exacerbates
the complexity of thermometry \cite{11,12}. For this reason, one establishes
the quantum thermometry theory based on the quantum parameter estimation to
finish this challenging task \cite{1,13,14,15,16,17,18}.

\bigskip The main goal of the quantum thermometry is to fulfil
high-precision temperature measurement by designing optimal thermometry
strategy in the quantum setting \cite{11,19,20}. To distinguish between
excellent and poor thermometry strategies, one usually resorts to the local
thermometry theory, which invokes the classical and quantum Cram\'{e}r-Rao
bound (CRB) based on unbiased estimators as indicators of optimal
thermometry precision \cite{1,10,21,22,23,24}. Such a thermometry theory
mainly focus on detecting tiny temperature variations around a known
temperature value \cite{8,11,25}. Nevertheless, one may not accurately know
the specific temperature value before measuring in the actual experimental
regimes \cite{25,26,27,28}. Moreover, the thermometry precision given by the
local thermometry theory is asymptotically tight only at the limit of
infinite multiple measurements, which may hardly provide the attainable
temperature precision limit with a limited number of measurements \cite%
{29,30,31,32,33}. Hence, one need to employ global thermometry theory based
on the bayesian estimation to tackle these problems \cite{25,26,34,35}. This
thermometry theory has attracted widely attention and produced many
important results in recent years \cite{8,11,25,26,27,28}. In particular, Jes%
\'{u}s Rubio \emph{et al}. proposed a novel approach for global quantum
thermometry according to the bayesian estimation theory \cite{25}. They
deemed that the quantum thermometry is a translation invariant estimation
problem, which leads to the fact that the mean logarithmic error (MLE) is
used for quantifying the estimation precision of the temperature instead of
the mean-square estimation error (MSE). Further, they derived a classical
lower bound for the mean logarithmic error via optimization estimator and
demonstrated that the indicators of the local thermometry theory hardly
apply to situations with limited data or insufficient prior information, in
which case the global thermometry is actually a good choice. Subsequently,
Mehboudi \emph{et al}. put forward a new ultimate bound for the MLE based on
the bayesian theory and constructed an adaptive scheme which can saturate
the ultimate bound \cite{11}. As a matter of fact, the global quantum
thermometry has been experimentally conducted on a specific experimental
equipment which is called release-recapture measurements on few tightly
confined $^{41}K$ atoms at $\mu K$ temperatures \cite{36}. However, these
research results mainly use some classical lower bounds for the MLE to study
the global quantum thermometry problems, without considering the
positive-operator-valued measure (POVM) \cite{37}. In other words, how to
derive a quantum lower bound for the MLE based on the POVM and further
investigate the quantum thermometry is still an open question.

On the other hand, the classical optimal biased bound (COBB) and quantum
optimal biased bound (QOBB) for the MSE is one of lower bounds which can
show better estimation performance of unknown parameters than conventional
classical and quantum CRB \cite{38,39,40}. We also noticed that the quantum
thermometry based on the COBB and QOBB has never been studied before.
Therefore, in this paper, we shall derive two ultimate bounds of the MLE on
thermometry precision in the global regime, i.e., COBB and QOBB. As an
illustration of this thermometry theory, we utilize these two lower bounds
as indicators for quantifying the temperature estimation precision for two
specific applications, i.e., noninteracting spin-$1/2$ gas \cite{41,42,43,44}
and a general $N$-level thermal equilibrium quantum probe \cite%
{1,45,46,47,48}. The results show that for these two specific applications,
the global quantum thermometry theory based on the COBB and QOBB can bespeak
better temperature estimation performance compared with the local one.

This paper is arranged as follows. In Sec. II, we briefly introduce the MLE
and the bias of the logarithm of the estimator. In Sec. III, we focus on the
general derivation of the COBB and QOBB for the MLE. In Sec. IV, we consider
two specific applications of quantum thermometry, i.e., noninteracting spin-$%
1/2$ gas and a general $N$-level thermal equilibrium quantum probe. Finally,
the main conclusions are drawn in the last section.

\section{Mean logarithmic error and the bias of the logarithm of the
estimator}

\bigskip Let us consider a physical system in thermal equilibrium whose the
temperature $T$ determined but unknown. In order to estimate the temperature
of the system, one usually implement the energy measurement on this system
to extract the information about the temperature and obtain the
corresponding conditional observation probability density $p(\varepsilon
|\theta ).$ It is worth noting that $\theta \in \lbrack \theta _{1},\theta
_{2}]$ denotes a hypothesis about the truth value of temperature rather than
true temperature for the global quantum thermometry \cite{25}. Further
assume that $p(\theta )$ is the prior probability density which
characterizes the known information about temperature prior to performing
energy measurements. According to Ref. \cite{25,49,50,51,52}, the
temperature is a scale parameter and thus the prior probability density $%
p(\theta )\varpropto 1/\theta $ is to satisfy scale invariance, i.e., \{$%
\varepsilon \mapsto \varepsilon ^{\prime }=\lambda \varepsilon ,\theta
\longmapsto \theta ^{\prime }=$ $\lambda \theta $\}, with $\lambda $
representing a dimensionless constant.

\bigskip For the sake of accurately evaluating the temperature of the system
from experimental measurement outcomes, one can resort to the MLE, which is
natural and suitable choice for the thermal equilibrium system.

\begin{equation}
MLE\equiv \int p(\varepsilon ,\theta )\left[ \ln \check{\theta}(\varepsilon
)-\ln \theta \right] ^{2}d\varepsilon d\theta ,  \label{1}
\end{equation}%
where $p(\varepsilon ,\theta )=p(\varepsilon |\theta )p(\theta )$ is the
joint probability density, $\check{\theta}(\varepsilon )$ is a point
estimator, which serves to map energy measurement results to temperature
values$.$ See Refs. \cite{8,25} for more detailed derivations and deeper
analysis.

\bigskip In an effort to conveniently find lower bounds for the MLE, we also
need to introduce the bias of the logarithm of the estimator%
\begin{equation}
b(\theta )=\int p(\varepsilon |\theta )\left[ \ln \check{\theta}(\varepsilon
)-\ln \theta \right] d\varepsilon ,  \label{2}
\end{equation}%
which is useful for the deduction of classical and quantum optimal biased
bounds for the MLE in the following section.

\section{\protect \bigskip Optimal biased bound for the mean logarithmic error%
}

In this section, our primary goal is to derive two lower bounds for the MLE,
i.e., COBB and QOBB based on the Cauchy-Schwarz inequality and the
variational method. These two types of lower bounds can respectively provide
the classical and quantum temperature estimation precision limit. Their
specific applications can be seen in the following section.

\subsection{\protect \bigskip Classical optimal biased bound}

\bigskip In this subsection, we shall derive a classical lower bound for the
MLE, i.e., COBB. For this reason, we introduce the average value of the
logarithm of the estimator with respect to the condition observation
probability density $p(\varepsilon |\theta ),$%
\begin{eqnarray}
A_{c}(\theta ) &=&\int p(\varepsilon |\theta )\ln \check{\theta}(\varepsilon
)d\varepsilon  \notag \\
&=&b(\theta )+\ln \theta .  \label{3}
\end{eqnarray}%
Thus, according to Eq. (\ref{3}), the corresponding MLE can also be
expressed as%
\begin{equation}
MLE=\int p(\theta )[Var(A_{c})+b(\theta )^{2}]d\theta ,  \label{4}
\end{equation}%
where $Var(A_{c})=\int p(\varepsilon |\theta )[\ln \check{\theta}%
(\varepsilon )-A_{c}(\theta )]^{2}d\varepsilon $ is the classical variance
of the logarithm of the estimator.

Note that Eq. (\ref{3}) has an equivalent form, i.e.,%
\begin{equation}
\int p(\varepsilon |\theta )[\ln \check{\theta}(\varepsilon )-A_{c}(\theta
)]d\varepsilon =0.  \label{5}
\end{equation}%
Differentiating the equivalent form Eq. (\ref{5}) with respect to $\theta ,$
one can obtain%
\begin{equation}
\int [\ln \check{\theta}(\varepsilon )-A_{c}(\theta )]p(\varepsilon |\theta )%
\frac{\partial \ln p(\varepsilon |\theta )}{\partial \theta }d\varepsilon
=A_{c}^{\prime }(\theta ),  \label{6}
\end{equation}%
where $A_{c}^{\prime }(\theta )=\partial A_{c}(\theta )/\partial \theta .$

\bigskip Multiplying both sides of Eq. (\ref{6}) by the prior probability
density $p(\theta )$ and a real function $z_{c}(\theta ),$ and then
integrating with respect to $\theta $, we find that%
\begin{equation}
\int xyd\varepsilon d\theta =\int A_{c}^{\prime }(\theta )p(\theta
)z_{c}(\theta )d\theta ,  \label{7}
\end{equation}%
where $x=\sqrt{p(\theta )p(\varepsilon |\theta )}[\ln \check{\theta}%
(\varepsilon )-A_{c}(\theta )]$ and $y=\sqrt{p(\theta )}z_{c}(\theta )\sqrt{%
p(\varepsilon |\theta )}\partial \ln p(\varepsilon |\theta )/\partial \theta
.$ According to the Cauchy-Schwarz inequality, we see that%
\begin{eqnarray}
&&\left \vert \int xyd\varepsilon d\theta \right \vert ^{2}  \notag \\
&\leq &\left( \int p(\theta )Var(A_{c})d\theta \right) \left( \int p(\theta
)z_{c}^{2}(\theta )F_{c}(\theta )d\theta \right) ,  \label{8}
\end{eqnarray}%
where $F_{c}(\theta )=\int p(\varepsilon |\theta )[\partial \ln
p(\varepsilon |\theta )/\partial \theta ]^{2}d\varepsilon $ is the classical
Fisher information.

\bigskip Based on Eqs. (\ref{7}) and (\ref{8}), we have%
\begin{equation}
\int p(\theta )Var(A_{c})d\theta \geq \frac{\left \vert \int A_{c}^{\prime
}(\theta )p(\theta )z_{c}(\theta )d\theta \right \vert ^{2}}{\int p(\theta
)z_{c}^{2}(\theta )F_{c}(\theta )d\theta },  \label{9}
\end{equation}%
which is always effective for the arbitrary real function $z_{c}(\theta )$
that satisfies the inequality $\int p(\theta )z_{c}^{2}(\theta )F_{c}(\theta
)d\theta >0$. Assuming $z_{c}(\theta )=A_{c}^{\prime }(\theta )/F_{c}(\theta
),$ we can get%
\begin{equation}
\int p(\theta )Var(A_{c})d\theta \geq \int p(\theta )\frac{\left[ b^{\prime
}(\theta )+1/\theta \right] ^{2}}{F_{c}(\theta )}d\theta .  \label{10}
\end{equation}%
Therefore, according to Eq. (\ref{4}), we further consider repeating the
same energy measurement $v$ times and can finally derive a novel classical
lower bound for the MLE, which is expressed as%
\begin{equation}
MLE\geq \int p(\theta )[b(\theta )^{2}+(b^{\prime }(\theta )+1/\theta
)^{2}/vF_{c}(\theta )]d\theta .  \label{11}
\end{equation}%
In order to minimize the lower bound given in Eq. (\ref{11}), we now follow
a procedure similar to the Ref. \cite{38,39} which derives the COBB for the
mean-square estimation error. Likewise, we invoke the variational principle
to find the optimal bias of the logarithm of the estimator $b(\theta )$
which minimizes the bound given in Eq. (\ref{11}). More specifically, let
the prior probability density $p(\theta )$ is at the rang of $a_{1}\leq $ $%
\theta $ $\leq a_{2}$ and define $\Gamma (b,\theta )=p(\theta )[b(\theta
)^{2}+(b^{\prime }(\theta )+1/\theta )^{2}/vF_{c}(\theta )],$ the optimal $%
b(\theta )$ can minimize $\int_{a_{2}}^{a_{1}}\Gamma (b,\theta )d\theta $ in
this situation and satisfy the Euler--Lagrange equation, i.e.,%
\begin{equation}
\frac{\partial \Gamma (b,\theta )}{\partial b(\theta )}-\frac{\partial }{%
\partial \theta }\frac{\partial \Gamma (b,\theta )}{\partial b^{\prime
}(\theta )}=0,  \label{12}
\end{equation}%
and the Neumann boundary condition%
\begin{equation}
\left. \frac{\partial \Gamma (b,\theta )}{\partial b^{\prime }(\theta )}%
\right \vert _{\theta =a_{1}}=\left. \frac{\partial \Gamma (b,\theta )}{%
\partial b^{\prime }(\theta )}\right \vert _{\theta =a_{2}}=0.  \label{13}
\end{equation}%
According to Eqs. (\ref{12}) and (\ref{13}), one can finally obtain the
differential equation for the optimal $b(\theta )$%
\begin{eqnarray}
&&b^{\prime \prime }(\theta )+\left[ \frac{p^{\prime }(\theta )}{p(\theta )}-%
\frac{F_{c}^{\prime }(\theta )}{F_{c}(\theta )}\right] b^{\prime }(\theta
)-vF_{c}(\theta )b(\theta )  \notag \\
&=&\frac{F_{c}^{\prime }(\theta )}{\theta F_{c}(\theta )}-\frac{p^{\prime
}(\theta )}{\theta p(\theta )}+\frac{1}{\theta ^{2}},  \label{14}
\end{eqnarray}%
with the boundary conditions $b^{\prime }(a_{1})=-1/a_{1}$ and $b^{\prime
}(a_{2})=-1/a_{2}.$ By solving the differential equation, one can get the
optimal $b(\theta ).$ Then, one can eventually derive the COBB for the MLE
by substitute the optimal $b(\theta )$ into Eq. (\ref{11}).

Before the end of this part, we consider a special case, i.e., recovery of
local thermometry. Without loss of generality, we follow Ref. \cite{25} and
posit that the temperature $T$ is considered fully unknown. Thus, one may
put into a more localized prior probability density $p(\theta )$ from Eq. (%
\ref{1}) \cite{25}. At this regime, the\ hypothesis $\theta $ shall be
mainly located within a narrow range, and the corresponding estimator $%
\check{\theta}(\varepsilon )$ shall approach to $\theta .$ Then, we can have
$\left[ \ln \check{\theta}(\varepsilon )-\ln \theta \right] ^{2}\approx %
\left[ (\check{\theta}(\varepsilon )-\theta )/\theta \right] ^{2},$ so that
Eq. (\ref{1}) can be reduced to%
\begin{equation}
MLE\approx \int p(\theta )\frac{Var[\check{\theta}(\varepsilon )]}{\theta
^{2}}d\theta ,  \label{15}
\end{equation}%
where $Var[\check{\theta}(\varepsilon )]=\int p(\varepsilon |\theta )[\check{%
\theta}(\varepsilon )-\theta ]^{2}d\theta $ is the classical variance of the
estimator $\check{\theta}(\varepsilon ).$ For an unbiased estimator, we
implement the same energy measurement $v$ times and have $Var[\check{\theta}%
(\varepsilon )]$ $\geq 1/vF_{c}(\theta ).$ Substituting the inequality into
Eq. (\ref{15}), we can get the classical Cram\'{e}r-Rao-like bound \cite{25}
\begin{equation}
MLE\gtrsim \int \frac{p(\theta )}{\theta ^{2}vF_{c}(\theta )}\equiv CCRLB.
\label{16}
\end{equation}%
It is noteworthy that Eq. (\ref{16}) is a local thermometry form as a limit
of the global thermometry \cite{25}.

\subsection{\protect \bigskip Quantum optimal biased bound}

Next, we will give a quantum lower bound for the MLE, i.e., QOBB. For the
quantum parameter estimation problem, any measurement can be expressed by a
POVM. Let $\hat{\rho}_{\theta }$ denote the density operator encoding the
parameter $\theta ,$ and then the conditional observation probability
density $p(\varepsilon |\theta )$ can be given by Born's rule \cite{37,53}
\begin{equation}
p(\varepsilon |\theta )=\text{Tr}(\hat{\rho}_{\theta }\hat{\Pi}_{\varepsilon
}),  \label{17}
\end{equation}%
\ where $\hat{\Pi}_{\varepsilon }$ is the POVM element related to the energy
measurement outcome. The corresponding average value of the logarithm of the
estimator with respect to the condition observation probability density $%
p(\varepsilon |\theta )$ can be rewritten as%
\begin{eqnarray}
A_{q}(\theta ) &=&\int \text{Tr}(\hat{\rho}_{\theta }\hat{\Pi}_{\varepsilon
})\ln \check{\theta}(\varepsilon )d\varepsilon  \notag \\
&=&b_{q}(\theta )+\ln \theta .  \label{18}
\end{eqnarray}%
After that, based on the Eq. (\ref{18}), the corresponding MLE is also
rephrased as%
\begin{eqnarray}
MLE &=&\int p(\theta )\text{Tr}(\hat{\rho}_{\theta }\hat{\Pi}_{\varepsilon })%
\left[ \ln \check{\theta}(\varepsilon )-\ln \theta \right] ^{2}d\varepsilon
d\theta  \notag \\
&=&\int p(\theta )[Var(A_{q})+b_{q}(\theta )^{2}]d\theta ,  \label{19}
\end{eqnarray}%
where $Var(A_{q})=\int $Tr$(\hat{\rho}_{\theta }\hat{\Pi}_{\varepsilon
})[\ln \check{\theta}(\varepsilon )-A_{q}(\theta )]^{2}d\varepsilon $ is the
quantum variance of the logarithm of the estimator.

\bigskip Likewise, there also exists an equivalent form for Eq. (\ref{18}),
i.e.,%
\begin{equation}
\int \text{Tr}(\hat{\rho}_{\theta }\hat{\Pi}_{\varepsilon })[\ln \check{%
\theta}(\varepsilon )-A_{q}(\theta )]d\varepsilon =0.  \label{20}
\end{equation}%
Further, differentiating Eq. (\ref{20}) with respect to $\theta ,$ we can get%
\begin{equation}
\int \text{Tr}(\hat{\Pi}_{\varepsilon }\partial \hat{\rho}_{\theta
}/\partial \theta )[\ln \check{\theta}(\varepsilon )-A_{q}(\theta
)]d\varepsilon =A_{q}^{\prime }(\theta ).  \label{21}
\end{equation}%
Multiplying both sides of Eq. (\ref{21}) by the prior probability density $%
p(\theta )$ and using the relation%
\begin{equation}
\frac{\partial \hat{\rho}_{\theta }}{\partial \theta }=\frac{1}{2}(\hat{\rho}%
_{\theta }\hat{L}_{\theta }+\hat{L}_{\theta }\hat{\rho}_{\theta }),
\label{22}
\end{equation}%
where $\hat{L}_{\theta }$ is the symmetric logarithmic derivative, we can
obtain%
\begin{eqnarray}
&&\text{Re}\int p(\theta )Tr(\hat{\rho}_{\theta }\hat{L}_{\theta }\hat{\Pi}%
_{\varepsilon })[\ln \check{\theta}(\varepsilon )-A_{q}(\theta )]d\varepsilon
\notag \\
&=&p(\theta )A_{q}^{\prime }(\theta ),  \label{23}
\end{eqnarray}%
where Re denotes the real part. Then, by multipling both sides of Eq. (\ref%
{23}) with a real function $z_{q}(\theta ),$ and then integrating with
respect to $\theta $, we see that%
\begin{equation}
\text{Re}\int \text{Tr}(\hat{X}^{\dagger }\hat{Y})d\varepsilon d\theta =\int
p(\theta )A_{q}^{\prime }(\theta )z_{q}(\theta )d\theta ,  \label{24}
\end{equation}%
where $\hat{X}^{\dagger }=\sqrt{p(\theta )}[\ln \check{\theta}(\varepsilon
)-A_{q}(\theta )]\sqrt{\hat{\Pi}_{\varepsilon }}\sqrt{\hat{\rho}_{\theta }}$
and $\hat{Y}=\sqrt{p(\theta )}z_{q}(\theta )\sqrt{\hat{\rho}_{\theta }}\hat{L%
}_{\theta }\sqrt{\hat{\Pi}_{\varepsilon }}.$ By invoking the Cauchy-Schwarz
inequality, we have
\begin{eqnarray}
&&\left \vert \text{Re}\int \text{Tr}(\hat{X}^{\dagger }\hat{Y})d\varepsilon
d\theta \right \vert ^{2}  \notag \\
&\leq &\left( \int p(\theta )Var(A_{q})d\theta \right) \left( \int p(\theta
)z_{q}^{2}(\theta )F_{q}(\theta )d\theta \right) ,  \label{25}
\end{eqnarray}%
where $F_{q}(\theta )=$Tr$(\hat{\rho}_{\theta }\hat{L}_{\theta }^{2})$ is
the quantum Fisher information.

\bigskip According to Eqs. (\ref{24}) and (\ref{25}), we can find
\begin{equation}
\int p(\theta )Var(A_{q})d\theta \geq \frac{\left \vert \int p(\theta
)A_{q}^{\prime }(\theta )z_{q}(\theta )d\theta \right \vert ^{2}}{\int
p(\theta )z_{q}^{2}(\theta )F_{q}(\theta )d\theta },  \label{26}
\end{equation}%
which is effective for any real function $z_{q}(\theta )$ that satisfies the
inequality $\int p(\theta )z_{q}^{2}(\theta )F_{q}(\theta )d\theta >0.$ In
particular, when postulating that $z_{q}(\theta )=A_{q}^{\prime }(\theta
)/F_{q}(\theta ),$ we get%
\begin{equation}
\int p(\theta )Var(A_{q})d\theta \geq \int p(\theta )\frac{\left[
b_{q}^{\prime }(\theta )+1/\theta \right] ^{2}}{F_{q}(\theta )}d\theta .
\label{27}
\end{equation}%
Thus, using the Eq. (\ref{19}), we further consider repeating the same
energy measurement $v$ times and can ultimately obtain a novel quantum lower
bound for the MLE, which is given by%
\begin{equation}
MLE\geq \int p(\theta )[b_{q}(\theta )^{2}+(b_{q}^{\prime }(\theta
)+1/\theta )^{2}/vF_{q}(\theta )]d\theta .  \label{28}
\end{equation}%
Similarly, we also need to seek the optimal bias of the logarithm of the
estimator $b_{q}(\theta )$ which can minimize the lower bound given in Eq. (%
\ref{28}). For this purpose, we will once again utilize the variational
principle to solve this problem. To be more specific, we posit the prior
probability density $p(\theta )$ is in $a_{1}\leq $ $\theta $ $\leq a_{2}$
and define $\Delta (b_{q},\theta )=p(\theta )[b_{q}(\theta
)^{2}+(b_{q}^{\prime }(\theta )+1/\theta )^{2}/vF_{q}(\theta )].$ At this
regime, the optimal $b_{q}(\theta )$ can minimize $\int_{a_{2}}^{a_{1}}%
\Delta (b_{q},\theta )d\theta $ in this situation and satisfy the
Euler--Lagrange equation, i.e.,%
\begin{equation}
\frac{\partial \Delta (b_{q},\theta )}{\partial b_{q}(\theta )}-\frac{%
\partial }{\partial \theta }\frac{\partial \Delta (b_{q},\theta )}{\partial
b_{q}^{\prime }(\theta )}=0,  \label{29}
\end{equation}%
with the Neumann boundary condition $\left. \partial \Delta (b_{q},\theta
)/\partial b_{q}^{\prime }(\theta )\right \vert _{\theta =a_{1}}=\left.
\partial \Delta (b_{q},\theta )/\partial b_{q}^{\prime }(\theta
)\right
\vert _{\theta =a_{2}}=0,$ so that%
\begin{eqnarray}
&&b_{q}^{\prime \prime }(\theta )+\left[ \frac{p^{\prime }(\theta )}{%
p(\theta )}-\frac{F_{q}^{\prime }(\theta )}{F_{q}(\theta )}\right]
b_{q}^{\prime }(\theta )-vF_{q}(\theta )b_{q}(\theta )  \notag \\
&=&\frac{F_{q}^{\prime }(\theta )}{\theta F_{q}(\theta )}-\frac{p^{\prime
}(\theta )}{\theta p(\theta )}+\frac{1}{\theta ^{2}},  \label{30}
\end{eqnarray}%
with the boundary conditions $b_{q}^{\prime }(a_{1})=-1/a_{1}$ and $%
b_{q}^{\prime }(a_{2})=-1/a_{2}.$ By solving the Eq. (\ref{30}), we can
obtain the corresponding optimal $b_{q}(\theta ).$ Substitute the optimal $%
b_{q}(\theta )$ into Eq. (\ref{28}), we can finally derive the QOBB for the
MLE.

\bigskip Likewise, the global quantum thermometry also has its local
thermometry form. In a similar way to derive Eq. (\ref{16}), one can obtain
the local thermometry form based on the Eq. (\ref{19}), i.e., quantum Cram\'{%
e}r-Rao-like bound
\begin{equation}
MLE\gtrsim \int \frac{p(\theta )}{\theta ^{2}vF_{q}(\theta )}\equiv QCRLB.
\label{31}
\end{equation}%
\begin{figure}[tbp]
\label{Fig1} \centering \includegraphics[width=0.9\columnwidth]{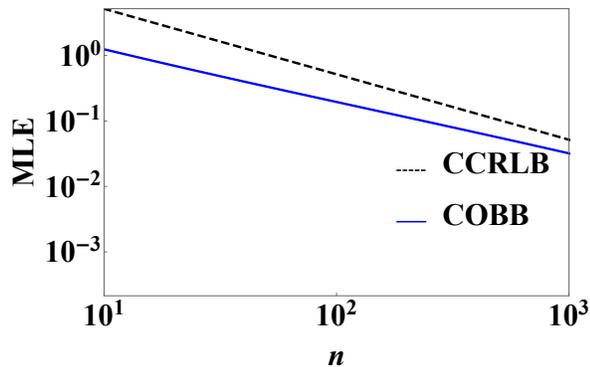}
\caption{{}(Color online) Log-log plot of the COBB and CCRLB as a function
of as a function of $n$. }
\end{figure}

\section{Applications}

\bigskip In order to intuitively show the estimation performance of the COBB
and QOBB, we consider two specific applications of quantum thermometry,
i.e., noninteracting spin-$1/2$ gas and a general $N$-level thermal
equilibrium quantum probe.

\subsection{Noninteracting spin-$1/2$ gas}

In this subsection, we consider a gas of $n$ noninteracting spin-$1/2$
particles with identical energy gaps $\varepsilon $ in thermal equilibrium.
In the following discussions, we set $\hbar =k_{B}=1.$ The conditional
observation probability density of measuring the total energy $%
E=r\varepsilon $ with $r\in \{0,1,...,n\}$ can be given by%
\begin{equation}
p(r|\theta )=\left(
\begin{array}{c}
n \\
r%
\end{array}%
\right) \frac{\exp (-r\varepsilon /\theta )}{Z(\varepsilon /\theta )},
\label{32}
\end{equation}%
where $Z(\varepsilon /\theta )=[1+\exp (-\varepsilon /\theta )]^{n}$ is the
partition function. Under the asymptotic condition of $n\gg 1,$ one can
derive the corresponding classical Fisher information \cite{25}%
\begin{eqnarray}
F_{c}(\theta ) &=&\sum \limits_{r}p(r|\theta )\left[ \frac{\partial }{%
\partial \theta }\ln p(r|\theta )\right] ^{2}  \notag \\
&=&\frac{n\varepsilon ^{2}/\theta ^{4}}{4\cosh ^{2}[\varepsilon /(2\theta )]}%
.  \label{33}
\end{eqnarray}%
Further presuppose the prior probability density $p(\theta )$ is in $0.1\leq
$ $\theta $ $\leq 10$ and set $\varepsilon =1,$ we can get the normalized
prior probability density $p(\theta )=1/(2\theta \ln 10).$ Substituting the
normalized prior probability density and Eq. (\ref{33}) into the Eqs. (\ref%
{11}), (\ref{14}) and (\ref{16}), we can respectively obtain the COBB and
CCRLB. For the sake of clearly seeing the estimation performance of the COBB
and CCRLB, we plot these two error bounds as a function of $n,$ as shown in
Fig. 1. It is clear that the performance of the COBB is always better than
the CCRLB.
\begin{figure}[tbp]
\label{Fig2} \centering \includegraphics[width=0.85\columnwidth]{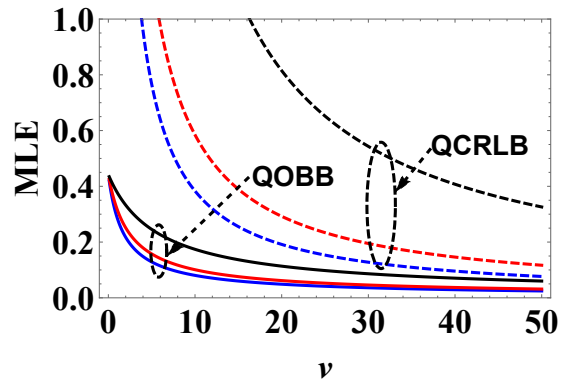}
\caption{{}(Color online) The QOBB and QCRLB as a function of the number of
energy measurements $v.$ The black, red and blue lines respectively
correspond to $N=2$, $N=4,$ and $N=6$. The dashed and solid lines correspond
to the QCRLB and QOBB, respectively.}
\end{figure}

\subsection{A general $N$-level thermal equilibrium quantum probe}

Then, we consider that a tiny thermometer interacts weakly with the sample
to be estimated and reaches the thermal equilibrium, so that the thermometer
is fully thermalized. The corresponding sample temperature can be inferred
from the probe state of the thermometer. We further presume that the
thermometer is in a general $N$-level thermal equilibrium quantum probe
state. According to Ref. \cite{1,26,45}, the quantum Fisher information of
the sample temperature hinges on the actual energy spectrum of the quantum
probe state. Correa \emph{et al}. showed that the optimal energy spectrum of
the general $N$-level probe state is a valid two-level system with a single
ground state and a $(N-1)$-fold degenerate excited state \cite{45}. They
also derived the maximum quantum Fisher information
\begin{equation}
F_{q}(\theta )=\frac{(N-1)\varepsilon ^{2}e^{\varepsilon /\theta }}{%
(N-1+e^{\varepsilon /\theta })^{2}\theta ^{4}},  \label{34}
\end{equation}%
where $\varepsilon $ is the energy gap between the ground and excited state.
We further assume that the prior probability density $p(\theta )$ is at the
rang of $0.1\leq $ $\theta $ $\leq 1,$ and set $\varepsilon =1,$ so that $%
p(\theta )=1/(\theta \ln 10).$ Substituting the prior probability density
and Eq. (\ref{34}) into the Eqs. (\ref{28}), (\ref{30}) and (\ref{31}), we
can respectively obtain the QOBB and QCRLB. In order to visually see the
estimation performance of the QOBB and QCRLB, we plot these two error bounds
as a function of the number of energy measurements $v$ for different values
of $N,$ involving $N=2$ (black lines), $N=4$ (red lines), and $N=6$ (blue
lines), as shown in Fig. 2. It is seen that for the QOBB and QCRLB, the
precision in temperature estimation can be effectively enhanced by
increasing the level $N$ of the probe state. Moreover, the estimation
performance of QOBB always outperforms the QCRLB.

\section{Conclusions}

In summary, we have used the variational method and the Cauchy-Schwarz
inequality to derive the classical and quantum versions of the optimal
biased bound for the MLE, i.e., COBB and QOBB. Our research is mainly based
on the global quantum thermometry techniques and makes use of these two
lower bounds as benchmarks for quantifying the temperature estimation
precision. Further, we have also analyzed the performance of these two lower
bounds for the temperature estimation by two specific applications, i.e.,
noninteracting spin-$1/2$ gas and a general $N$-level thermal equilibrium
quantum probe. First, we have investigated the temperature estimation
problem of the noninteracting spin-$1/2$ gas. The numerical results show
that compared with the CCRLB based on the local thermometry, the COBB based
on the global one can exhibit better estimation performance. Then, we have
also considered that a general $N$-level thermalized quantum probe
accurately evaluate the sample temperature. We find that similar to the
QCRLB, the temperature precision limit determined by the QOBB can be validly
improved by increasing the level $N$ of the probe state and the estimation
performance of QOBB consistently surpasses that of QCRLB.

\begin{acknowledgments}
This work was supported by the National Nature Science Foundation of China
(Grant Nos. 91536115, 11534008, 62161029); Natural Science Foundation of
Shaanxi Province (Grant No. 2016JM1005); Shaanxi Fundamental Science
Research Project of Mathematics and Physics (Grant No. 22JSY005); Natural
Science Foundation of Jiangxi Provincial (Grant No. 20202BABL202002). Wei Ye
is supported by the Scientific Research Startup Foundation (Grant No.
EA202204230) at Nanchang Hangkong University.
\end{acknowledgments}

\end{document}